\documentclass[11pt,twoside]{article}

\usepackage{galev06}
\usepackage{epsf}
\usepackage{psfig}
\usepackage{lscape}
\usepackage{graphicx}

\begin{document}
\title{Triggering and Tempering Star Formation in Colliding Galaxies}   
\author{Curtis Struck}   
\affil{Dept. of Physics and Astronomy, Iowa State Univ., Ames, IA 50011, USA}    

\begin{abstract} 
Recent observational literature suggests that only a small number of colliding galaxies show substantial star formation enhancements before merging. Most have SFRs comparable to those of late-type field galaxies. Preliminary Spitzer studies (with data from the Spirals, Bridges and Tails and SINGs projects) generally confirm this, though with some caveats. If star formation in isolated disks is self-regulated, while collisions disequilibrate much of the disk, then the similarity to star-formation in isolated disks is surprising. Comparisons of observations and dynamical models suggest some resolutions to this paradox, including the role of downsizing. The interesting example of the Arp 82 system is presented.
\end{abstract}



When and how do early-stage colliding galaxies enhance star formation vs. rearranging or reducing (tempering) it, e.g., in new induced waves or bars? Recently, in part as an offshoot of large scale galaxy surveys, there have been a number of studies of star formation in galaxy groups and interacting systems, (see e.g., the review of \citet{str06} and references therein). Briefly, the general results of these works include the following: 1) there is star formation enhancement in early stage galaxy collisions but the amount is modest, 2) most of the enhancement is due to core starbursts, rather than extended star formation, 3) there is evidence for a proximity effect, i.e., more enhancement at smaller separations between the galaxies, 4) there is also evidence for a velocity proximity effect, i.e., more enhancement at smaller relative velocities. 

The Spitzer Cycle 1 'Spirals, Bridges and Tails' (SB\&T) project was designed to study the sites and modes of induced SF in a sample of quite strongly interacting, but pre-merger galaxies with IRAC and MIPS images. The sample contains about two dozen Arp atlas galaxies, which have been supplemented by more interacting systems taken from the Spitzer archive, to make a total sample of about three dozen galaxies. The object names and images can be viewed at the public website - http://www.etsu.edu/physics/bsmith/research/sbt.html. We note that this work is very much in the spirit of the earlier IRAS study of \citet{bus88}, but with the better resolution of the Spitzer instruments.  Complementary GALEX observations of a similarly sized and overlapping sample are underway, and H$\alpha$ observations of the combined samples are in progress. A second goal of the SB\&T project is the comparison of different star formation indicators in interacting systems. 

Here I report a number of preliminary SB\&T results. More detailed results are reported in \citet{smi06}. First, I should note that to judge star formation enhancements in our sample, we must have a control sample with which to compare. The Spitzer SINGS Legacy project provides a good number of relatively isolated galaxies, of a variety of Hubble types which serve that purpose, see \citet{ken03}. Note that in constructing our sample of 'isolated' SINGS spirals, we only included galaxies without nearby massive companions.

The left hand graph in Figure 1 shows the IRAC band 1 to band 2 color distributions for various SB\&T subsamples in comparison to the SINGS 'spirals.' There is little difference between the top four panels. We believe that in these galaxies the emission in these bands is dominated by stars with intermediate to old ages. Evidently,  there is little difference between the colors of these populations in isolated spirals, interacting disks, the disks of M51 type interactions,  in tidal bridges and tails, or even in ellipticals. 

\begin{figure}[!h]
\includegraphics[scale=0.33]{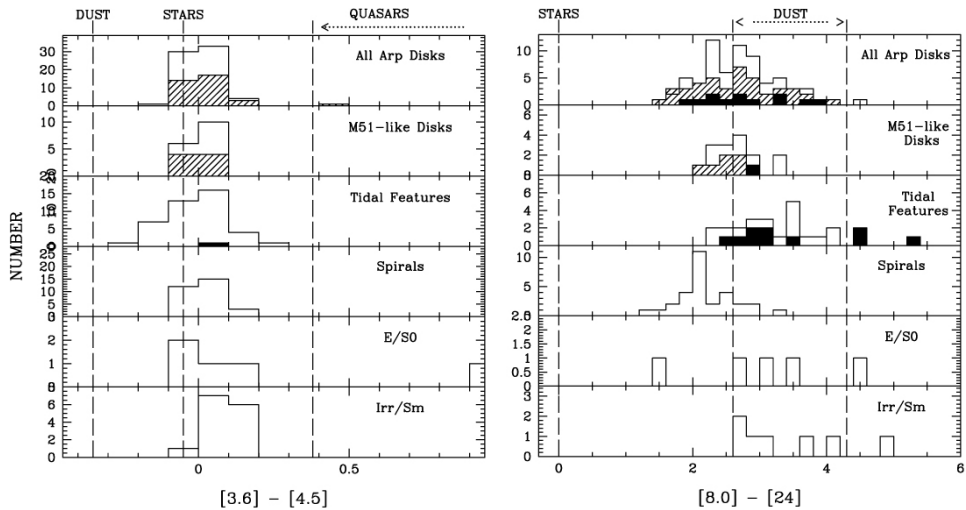} 
 \caption{Left graph; histograms of the Spitzer IRAC [3.6] - [4.5] colors for the different samples: the Arp galaxy main disks (top panel), the subset of M51-like systems (second panel), tidal features (third panel), the spiral galaxies (fourth panel), ellipticals/S0 (fifth panel), and irregular/Sm galaxies (bottom panel). The hatched galaxies are the more massive in the pair. The mean [3.6] - [4.5] color for the \citet{whi04} field stars is -0.05, while the predicted value for interstellar dust is -0.35 \citep{li01}. The Arp disk with the very red color of 0.44 is the Seyfert NGC 7469 (Arp 298S), while the red elliptical is NGC 1377. Right graph:histograms of the Spitzer IRAC [8.0] - MIPS [24] colors for the different samples. The filled areas represent upper limits. The hatched galaxies are the more massive in the pair. The expected [8.0] - [24] color for dust varies from 2.6 - 4.3 \citep{li01}, increasing with increasing interstellar radiation field intensity, while stars are expected to be at about 0.0. Graphs from \citet{smi06}.}
 \end{figure}

The right hand graph in Figure 1 shows the IRAC band 4 (8.0 micron) to MIPS 24 micron color distributions. In this figure there is a clear and statistically significant difference between the isolated spirals and the interaction components. The latter distribution has a significant red tail in this color, which in these bands suggests more emission from dust heated by embedded young stellar populations. This result is supported by the fact that the IRAS fluxes of the interacting sample are about a factor of 2 larger (per galaxy) than those of the comparison sample, in agreement with the results of \citet{bus88} on a similar sample.

Many of the systems with the reddest colors show evidence for core  starbursts as also found in optical surveys. In any case, the enhancements we see confirm the importance of dynamical triggering from the time of the first close pass, at least in the cores. 

On the other hand, we do not see any evidence of a proximity effect in this sample.  This may be due to the small sample size and the fact that the galaxies were selected to be strongly tidally distorted. In the group surveys with large samples the proximity effect is most prominent among subsamples with high star formation rates, when the separation is less than about 25 kpc, and when there is also a velocity proximity effect. This combination suggests that the effect might be the result of the onset of merging in a second or final encounter. There are few such systems in the SB\&T sample. 

The rapid onset of such triggering precludes some explanations of how it is achieved, for example, via the development of a bar, and bar-driven funneling of gas to the nucleus. (Moreover, there are not many strongly barred systems in the sample.) The systems with apparent core starbursts include  a number of the smaller companions, especially in the M51 type systems. These companions have probably been strongly disturbed in the most recent encounter, and have experienced mass accretion via a bridge. Numerical models suggest that it takes some time for much of this material to fall onto the companion, and that the peak of induced star formation may occur near apogalacticon, when the infalling gas can catch up to the companion, rather than at closest approach. Other systems with core starbursts include ring galaxies, and the few systems in our sample with strong bars. 

Each of these early stage interacting systems has a distinct dynamical story of its own. There is much to be learned from these stories, though unraveling them requires multi-waveband observations, and specifically tailored numerical models.  We have been able to carry out such studies on only a few of the SB\&T systems so far. 

One very interesting example is the Arp 82 (NGC~2535/6) system. This M51 type encounter was previously investigated in some detail by \citet{kau97}. The primary has a long tidal tail, a substantial bridge, and an 'ocular' (eye-shaped) waveform containing knots of star formation. Despite these strong tidal morphologies, the Spitzer colors alone do not indicate especially strong recent star formation relative to SINGS comparison galaxies, however, the compact companion does contain a starburst. 

The mystery deepens when we recall that Kaufman et al.\ found that the system contains an unusually high amount of HI gas (with little molecular gas), i.e. the HI mass almost equals the estimated stellar mass.  Moreover, Hancock has fitted Starburst99 population models to new GALEX and optical data, and found moderate extinctions and young ages for the clump sources \citep{han06}. There was no evidence for a stellar population older than about 400 Myr. in the diffuse emission. The combination of young and intermediate populations can explain the normal disk Spitzer colors. 

It appears that if the encounter has been of extended duration, and most of the visible stars could have been formed in the interaction. A hydrodynamical model presented in Hancock et al.\ confirms that the time since the last closest approach is of this order, and the feedback prescription used in this model also confirms that much of the star formation could also be the result of the interaction. 

Evidently, the two progenitors were low surface brightness galaxies or of very late Hubble type, and quite unevolved before the interaction. The 3.6 $\mu$m luminosities of the two galaxies are low relative to the SINGS galaxies, so they are probably also less massive than ordinary spirals. Thus, they seem to provide a nice example of the mass downsizing phenomenon in action, and of the importance of intermediate-mass galaxies in the cosmic SF at the present time. 

Arp 82 is not unique as an interacting system with gas-rich, unevolved  progenitors; the Cartwheel ring galaxy is probably another example.  However, they are  not the majority in the SB\&T sample. Intermediate mass systems are common. This recalls the result of \citet{hin06} that a number of nearby LIRGs appear to be low mass, or downsized, ULIRGs.

In sum, optical/UV measures of star formation in colliding galaxies suffer extinction problems. Young populations can be masked in the near-IR by dominant old populations. Therefore, resolved mid and far-IR observations have great advantages. It appears that strong interactions enhance star formation by a factor of about 2-3 in the SB\&T sample, in agreement with earlier studies. The resolved Spitzer observations show the spatial distribution of the star formation, and as often as not, the starbursts are in the lower mass, but more disturbed companion. Diversity in star formation history seems to be the rule in early stage collisions, so averages of the enhancement or proximity effects only tell part of the story. In judging enhancement, one must be aware of the role of downsizing in present day interactions.  Most of the star formation enhancement in interacting systems occurs in small, highly disturbed companions or a few special types of interaction, with the star formation rate much less affected than the disk morphology in many other systems.

\acknowledgements I am very grateful to the members of the SB\&T and Ocular collaborations for teaching me much about these objects and observations of them.


\end{document}